\documentclass[preprint,showpacs,preprintnumbers,amsmath,amssymb]{revtex4-1}
\usepackage{latexsym}
\usepackage{epsfig}
\usepackage{float}
\usepackage{lipsum}
\usepackage{graphicx,times}
\usepackage{amssymb,amsmath}
\usepackage{color}
\usepackage{dcolumn}
\usepackage{epstopdf}
\usepackage{pgfplots}
\usepackage[colorlinks=true, urlcolor=blue, linkcolor=blue, citecolor=blue]{hyperref}
\newcommand{\be}{\begin{equation}}
\newcommand{\ee}{\end{equation}}
\begin{document}
\title{Structure, formation and decay of $\bar{K}NN$ system by Faddeev-AGS calculations}
\author{S. Marri}
\affiliation{Department of  Physics, Isfahan University of Technology, Isfahan 84156-83111, Iran}
\author{S. Z. Kalantari}
\affiliation{Department of  Physics, Isfahan University of Technology, Isfahan 84156-83111, Iran}
\author{J. Esmaili}
\affiliation{Department of Physics, Faculty of Basic Sciences, Shahrekord University, Shahrekord, 115, Iran}
\date{\today}
\begin{abstract}
The Faddeev AGS equations are solved for coupled-channels $\bar{K}NN-\pi\Sigma{N}$ system 
with quantum numbers $I=1/2$ and $S=0$. Using separable potentials for $\bar{K}N-\pi\Sigma$ 
interaction, we have calculated the transition probability for the $(Y_{K})_{I=0}+N\rightarrow\pi\Sigma{N}$ 
reaction. The possibility to observe the trace of $K^{-}pp$ quasi-bound state in the $\pi\Sigma{N}$ 
mass spectra was studied. Various types of chiral based and phenomenological potentials are 
used to describe the $\bar{K}N-\pi\Sigma$ interaction. It was shown that not only we can see the 
signature of the $K^{-}pp$ quasi-bound state in the mass spectra, but also, one can see the trace 
of branch points in the observables.
\end{abstract}
\maketitle
%%%%%%%%%%%%%%%%%%%%%%%%%%%%%%%%%%%%%%%%%%%%%%%%%%%%%%%%%%%%%%%%%%%%%%%%%%%%%%%%%%%%%%%%%%%%%%%%%%%
\section{Introduction}
\label{intro}
%%%%%%%%%%%%%%%%%%%%%%%%%%%%%%%%%%%%%%%%%%%%%%%%%%%%%%%%%%%%%%%%%%%%%%%%%%%%%%%%%%%%%%%%%%%%%%%%%%%
The wish to define a precise interaction model for $\bar{K}N$ interaction is a basic 
goal in strangeness nuclear physics. For the past two decades, an enormous amount of 
efforts has been made to study the structure of dense kaonic nuclear 
clusters~\cite{akaishi,yamazaki1,dot1,dot2}. An important kaonic cluster is the $K^{-}pp$ 
system, which is a highly controversies issue in studying the kaonic systems. Many 
theoretical calculations were performed, focusing on the $K^{-}pp$ 
system~\cite{shev1,shev2,ikeda1,ikeda2,dote1,dote2,ikeda3,kh1,kh2,kh3,maeda}.  All few-body 
calculations have shown that the $K^{-}pp$ is bound, but with some variation in the 
values of the extracted pole energy. 

\begin{figure}[H]
\vspace{0.0cm}
\centering
\includegraphics[scale=0.7]{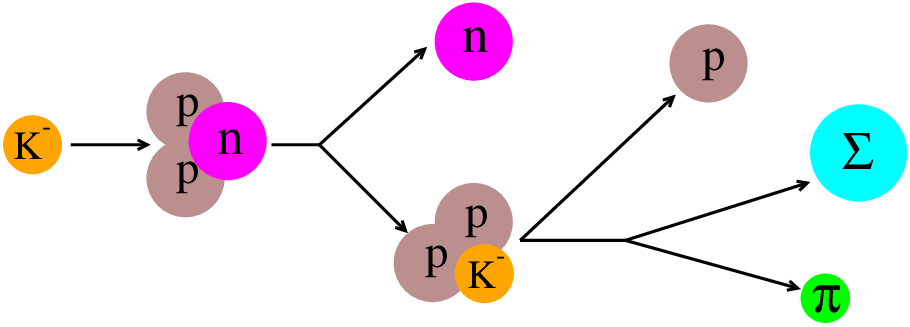}
\vspace{0.0cm}
\caption{Diagram for the $\mathrm{^{3}{He}}(K^{-},N)K^{-}pp$ reactions.}
\label{diag}
\end{figure}

If $K^{-}pp$ system is indeed bound, then the remaining question is whether this state 
is sufficiently narrow to allow observation and identification. Due to strong absorption 
of antikaon by the nucleus, the quasi-bound state in $\bar{K}NN$ system can have a 
large width. Thus, this may provide difficulties for direct experimental observation 
of kaonic bound states in nuclei~\cite{koike}. Many experimental efforts have been also 
performed to explore the pole structure of the $K^{-}pp$ system. An exclusive analysis of the 
$p+p\rightarrow{X}+K^{+},X\rightarrow{p}+\Lambda$ reaction at 2.85 GeV~\cite{yamazaki2} 
indicated a large peak both in the $\Lambda{p}$ invariant-mass and $K^{+}$ missing-mass 
spectra, which had been predicted in a theoretical works~\cite{yamazaki3,yamazaki4}. The 
observed peak corresponds to the binding energy of about 103 MeV and the width is given 
as $\Gamma=$118 MeV. The $K^{-}pp$ quasi-bound state can be produced in kaon-induced reactions 
on light nuclei such as $\mathrm{^{3}{He}}$ and deuteron, and the signal of the resonance 
may be observed in the mass spectra of the final particles. The investigation for the 
$K^{-}pp$ quasi-bound state have been further explored through $\pi^{+}$ incident reaction 
$d(\pi^{+},K^{+})K^{-}pp$ by E27 experiment at J-PARC~\cite{e27}. The $d(\pi^{+},K^{+})$ 
experiment also revealed a distinct peak in the $K^{+}$ missing-mass spectrum, nearly at 
the same mass and width as the DISTO peak X. The investigation for the $K^{-}pp$ quasi-bound 
state could be reached through the $K^{-}+\mathrm{^{3}{He}}$ reaction (see Fig.~\ref{diag}). 
This reaction was performed as an E15 experiment at J-PARC~\cite{e15}. The E15 group suggests 
a broad $K^{-}pp$ quasi-bound state structure at 15 MeV just below $\bar{K}NN$ 
threshold~\cite{e15}.

The kaon-induced reactions have been studied by Koike-Harada~\cite{harada} and Yamagata 
{\it et al.}~\cite{yamagata} using the optical potential method. In Ref.~\cite{ohnishi} 
the $(Y_{K})_{I=0}+N\rightarrow\pi\Sigma{N}$ has been studied by using Faddeev approach. 
In this calculation they employed chiral based potentials for the $s$-wave $\bar{K}N$ 
interaction. Within their model, they have found a clear signal corresponding to the 
strange-dibaryon resonances in the Faddeev scattering amplitudes and the $(Y_{K})_{I=0}+N\rightarrow\pi\Sigma{N}$ 
transition probabilities. The $K^{-}+\mathrm{^{3}{He}}\rightarrow\Lambda{pn}$ reaction 
was also studied by Sekihara {\it et al.},~\cite{sek1,sek2,sek3} to investigate the origin 
of observed peak close to the $K^{-}pp$ threshold in first run of E15 experiment at 
J-PARC~\cite{e15}. Two scenarios were considered to produce the peak. In the first scenario, 
the $\Lambda$(1405) resonance can be generated but it does not correlate with $p$, and the 
uncorrelated $\Lambda(1405)p$ system subsequently decays into $\Lambda{p}$ and in the other 
one, the $\bar{K}NN$ quasi-bound state should be generated and decays into $\Lambda{p}$. 
From the calculation of the $\Lambda{p}$ invariant mass spectrum, the experimental 
signal was reproduced in the second scenario and they definitely discarded the scenario 
that the $\Lambda$(1405) does not correlate with $p$.

%%%%%%%%%%%%%%%%%%%%%%%%%%%%%fig.1
%%%%%%%%%%%%%%%%%%%%%%%%%%%%%

This study is devoted to study the pole structure of the $K^{-}pp$ three-body system. We 
study how well the signal $K^{-}pp$ quasi-bound state can be observed in the $\pi\Sigma{N}$ 
mass spectra resulting from reaction under consideration. We performed few-body calculations 
for the $\bar{K}NN-\pi\Sigma{N}$ system by using coupled-channels Faddeev AGS equations. 
The transition probabilities for the $(Y_{K})_{I=0}+N\rightarrow\pi\Sigma{N}$ reaction are 
calculated. With this method, we investigated the behavior of the transition probability 
for $(Y_{K})_{I=0}+N\rightarrow\pi\Sigma{N}$ reaction. Several chiral based and phenomenological 
$\bar{K}N-\pi\Sigma$ potentials are used~\cite{shev3,shev4,ohnishi} to investigate the 
sensitivity of the the three-body observables on two-body inputs.

The paper is organized as follows: in Sect.~\ref{formula}, we will explain the formalism 
used for the coupled channel three-body $\bar{K}NN-\pi\Sigma{N}$ system and give a brief 
description of the transition probability formula for break-up reactions. Sect.~\ref{put} 
is devoted to the two-body inputs of the calculations. The computed transition probabilities 
are presented in Sect.~\ref{result} and in Section~\ref{conc}, we give conclusions.
%%%%%%%%%%%%%%%%%%%%%%%%%%%%%%%%%%%%%%%%%%%%%%%%%%%%%%%%%%%%%%%%%%%%%%%%%%%%%%%%%%%%%%%%%%%%%%%%%%%
\section{Three particle system $\bar{K}NN-\pi\Sigma{N}$}
\label{formula}
%%%%%%%%%%%%%%%%%%%%%%%%%%%%%%%%%%%%%%%%%%%%%%%%%%%%%%%%%%%%%%%%%%%%%%%%%%%%%%%%%%%%%%%%%%%%%%%%%%%
The calculation of the $\bar{K}NN-\pi\Sigma{N}$ three body system is based on the Faddeev 
treatment~\cite{alt}. Separable potentials were used for describing the two-body interactions
\begin{equation}
V_{i,I_{i}}^{\alpha\beta}(k^{\alpha}_{i},k^{\beta}_{i};z)=g_{i,I_{i}}^{\alpha}(k^{\alpha}_{i})
\lambda_{i,I_{i}}^{\alpha\beta}g_{i,I_{i}}^{\beta}(k^{\beta}_{i}),
\label{eq1}
\end{equation}
where 
$g_{i,I_{i}}^{\alpha}(k^{\alpha}_{i})$ is the form factor of the interacting two-body 
subsystem $(jk)$, with relative momentum $k^{\alpha}_{i}$ and isospin $I_{i}$. Here, 
$\lambda_{i,I_{i}}^{\alpha\beta}$ is the strength parameter of the interaction. To take 
the $\bar{K}N-\pi\Sigma$ coupling directly into account, the potentials are further 
labeled with the $\alpha$ values. The two-body $T$-matrices in separable form can be given by
\begin{equation}
\begin{split}
T_{i,I_{i}}^{\alpha\beta}(k^{\alpha}_{i},k^{\beta}_{i};z)=g_{i,I_{i}}^{\alpha}(k^{\alpha}_{i})
\tau_{i,I_{i}}^{\alpha\beta}(z-E^{\alpha}_{i}(p^{\alpha}_{i}))g_{i,I_i}^{\beta}(k^{\beta}_{i}),
\end{split}
\label{eq2}
\end{equation}
where $z$ and $E^{\alpha}_{i}(p^{\alpha}_{i})$ are the total energy of the system and the energy 
of the spectator particle in $\alpha$ channel, respectively. 
\begin{equation}
E^{\alpha}_{i}(p^{\alpha}_{i})=\frac{(p^{\alpha}_{i})^{2}}{2\nu^{\alpha}_{i}}, \hspace{0.5cm}
\label{eq3}
\end{equation}
the quantity $\nu^{\alpha}_{i}=m^{\alpha}_{i}(m^{\alpha}_{j}+m^{\alpha}_{k})/(m^{\alpha}_{i}
+m^{\alpha}_{j}+m^{\alpha}_{k})$, is the reduced mass, when particle $i$ in channel $\alpha$ 
is spectator. The operator $\tau_{i,I_{i}}^{\alpha\beta}(z-E^{\alpha}_{i}(p^{\alpha}_{i}))$ 
is also the usual two-body propagator. Using separable potential for two-body interactions, 
the three-body Faddeev equations~\cite{shev2} in the AGS take the form

\begin{equation}
\begin{split}
& \mathcal{K}_{ij,I_{i} I_{j}}^{\alpha\beta}(\vec{p}^{\,\alpha}_{i},\vec{p}^{\,\beta}_{j};z)= 
\delta_{\alpha\beta} \mathcal{M}_{ij,I_{i} I_{j}}^{\alpha\beta}(\vec{p}^{\,\alpha}_{i},\vec{p}^{\,\beta}_{j};z) \\
& +\sum_{k,I_{k},\gamma}\int{d}\vec{p}^{\,\alpha}_{k}
\mathcal{M}_{ik,I_i I_k}^\alpha(\vec{p}^{\,\alpha}_{i},\vec{p}^{\,\alpha}_{k};z)
\tau_{k,I_k}^{\alpha\gamma}(z-E^{\alpha}_{k}(\vec{p}^{\,\alpha}_{k})) \\
& \times\mathcal{K}_{kj,I_k I_j}^{\gamma\beta}(\vec{p}^{\,\gamma}_{k},\vec{p}^{\,\beta}_{j};z).
\end{split}
\label{eq444}
\end{equation}

After partial wave decomposition and where we assumed that only s-wave contribution will 
be significant in our calculations, we get the following equations

\begin{equation}
\begin{split}
& \mathcal{K}_{ij,I_{i} I_{j}}^{\alpha\beta}(p^{\alpha}_{i},p^{\beta}_{j};z)= \delta_{\alpha\beta}
\mathcal{M}_{ij,I_{i} I_{j}}^{\alpha\beta}(p^{\alpha}_{i},p^{\beta}_{j};z) \\
& +\sum_{k,I_{k},\gamma}\int{d}^{3}p^{\alpha}_{k}\mathcal{M}_{ik,I_i I_k}^\alpha(p^{\alpha}_{i},p^{\alpha}_{k};z)
\tau_{k,I_k}^{\alpha\gamma}(z-E^{\alpha}_{k}(p^{\alpha}_{k})) \\
& \times\mathcal{K}_{kj,I_k I_j}^{\gamma\beta}(p^{\gamma}_{k},p^{\beta}_{j};z).
\end{split}
\label{eq4}
\end{equation}

Here, the operators $\mathcal{K}_{ij,I_{i}I_{j}}^{\alpha\beta}$ are the transition amplitudes 
between Faddeev channels and particle channels~\cite{shev2}. The operators 
$\mathcal{M}_{ij,I_{i}I_{j}}^{\alpha\beta}$ are the corresponding Born terms. The inputs for the 
AGS system of equations (\ref{eq4}) are two-body $T$-matrices, embedded in the three-body Hilbert 
space. Faddeev partition indices $i,j,k=1,2,3$ are used to define the interacting pair and 
also the spectator particle. The Faddeev equations are modified~\cite{shev2,ikeda1} to take the 
$\bar{K}N-\pi\Sigma$ coupling directly into account. Thus, in addition to the Faddeev indices the 
particle indices ($\alpha,\beta,\gamma=1,2,3$) are also added for each state ($i$)~\cite{revei}.
$$
\alpha=\{1,2,3\}=\{\bar{K}NN,\pi\Sigma{N}_{2},\pi{N}_{1}\Sigma\}.
$$

Since the total isospin of the system is $I=1/2$, therefore, depending on the spin of the two baryons, 
we should treat $K^{-}pp$ or $K^{-}d$ system. The total spin of the system remains unchanged. Therefore, 
the baryon spins do not enter explicitly and the operators will be labeled by isospin indices. In the 
$K^{-}pp$ system the spin component is antisymmetric, so all operators in isospin base should be symmetric.

In present Calculations, we used the quasi-particle approach to solve the Faddeev equations for bound 
state problem. The most important part of the quasi-particle method is the separable expansion of the 
scattering amplitudes in the two- and three-body systems~\cite{grass,nadro,fonce}. To find the $K^{-}pp$ 
pole position, the separable representation must be defined for the three-body amplitudes and driving 
terms. For this purpose, we used the Hilbert-Schmidt expansion method~\cite{nadro,fix}.

\begin{equation}
\begin{split}
& \mathcal{M}_{ij,I_{i}I_{j}}^\alpha(p^{\alpha}_{i},p^{\alpha}_{j};z)=-\sum_{n=1}\lambda_n(z)
u_{n;i,I_i}^\alpha(p^{\alpha}_{i};z) \\
& \hspace{2.8cm}\times u_{n;j,I_j}^\alpha(p^{\alpha}_{j};z),
\end{split}
\label{eq5}
\end{equation}
the form factors $u_{n;i,I_i}^\alpha(p^{\alpha}_{i},z)$ are taken as the eigenfunctions of the kernel of 
the equation (\ref{eq4}). The separable form of the Faddeev transition amplitudes is given by
\begin{equation}
\mathcal{K}_{ij,I_i I_j}^{\alpha\beta}(p^{\alpha}_{i},p^{\beta}_{j};z)=\sum_{n}
u_{n;i,I_i}^\alpha(p^{\alpha}_{i};z)\zeta_{n}(z)
u_{n;j,I_j}^\beta(p^{\beta}_{j};z),
\label{eq6}
\end{equation}
where the functions $\zeta_{n}(z)$ obey the equation
\begin{equation}
\zeta_n(z)=\lambda_n(z)/(\lambda_n(z)-1).
\label{eq7}
\end{equation}

Applying Hilbert-Schmidt expansion method~\cite{fix} to the Faddeev equations of $\bar{K}NN-\pi\Sigma{N}$, 
the following homogeneous integral equations for $u_{n;i,I_i}^\alpha(p^{\alpha}_{i};z)$ are obtained
\begin{equation}
\begin{split}
& u_{n;i,I_i}^\alpha(p^{\alpha}_{i};z)=\frac{1}{\lambda_n}\sum_{\gamma,k=1}^{3}\sum_{I_k}
\int{d}^{3}p^{\alpha}_{k}
\mathcal{M}_{ik,I_i I_k}^\alpha(p^{\alpha}_{i},p^{\alpha}_{k};z) \\
& \hspace{1.9cm}\times\tau_{k,I_k}^{\alpha\gamma}(z-E^{\alpha}_{k}(p^{\alpha}_{k}))
u_{n;k,I_k}^\gamma(p^{\gamma}_{k};z).
\end{split}
\label{eq8}
\end{equation}

To solve the homogeneous system, we should search for a complex energy at which one of the eigenvalues 
($\lambda_{n}(z)$) of the kernel matrix becomes equal to one. We must work on the physical and unphysical 
energy sheet of the $\bar{K}NN$ and $\pi\Sigma{N}$ channels, respectively. 

Another purpose of this work is to study the possible signature of the $K^{-}pp$ quasi-bound state in 
the $\pi\Sigma{N}$ mass spectra from the reaction $(Y_{K})_{I=0}+N\rightarrow\pi\Sigma{N}$. The break-up 
amplitude for this reaction in terms of the Faddeev transition amplitudes can be given by~\cite{afnan}.
\begin{equation}
\begin{split}
& T_{\pi\Sigma{N}\leftarrow(Y_{K})_{I=0}+N} (\vec{k}_N,\vec{p}_N,p'_{N};z) \\
& =\sum_{I}g_{\pi\Sigma,I}(\vec{k}_N )\tau_{{(\pi\Sigma)}_I N,(\bar{K}N)_IN}(z-E_N(\vec{p}_N )) \\
& \hspace{1.cm}\times\mathcal{K}_{(\bar{K}N)_I N,(\bar{K}N)_{I=0} N}(p_{N},p'_{N};z) \\
& +\sum_{I}g_{\pi\Sigma,I}(\vec{k}_N )\tau_{(\pi\Sigma)_I N,(\pi\Sigma)_I N}(z-E_N(\vec{p}_N )) \\
& \hspace{1.cm}\times\mathcal{K}_{(\pi\Sigma)_I N,(\bar{K}N)_{I=0}N}(p_{N},p'_{N};z) \\
& +\sum_{I}\sum_{I'}\langle[\pi\otimes\Sigma]_{I'}\otimes{N}\mid\pi\otimes[\Sigma\otimes{N}]_{I}
\rangle{g}_{\Sigma N,I}(\vec{k}_\pi ) \\
& \hspace{1.cm}\times \tau_{\pi(\Sigma N)_I,\pi(\Sigma N)_I }(z-E_{\pi} (\vec{p}_\pi)) \\
& \hspace{1.cm}\times \mathcal{K}_{\pi(\Sigma N)_I,(\bar{K}N)_{I=0}N}(p_{\pi},p'_{N};z),
\end{split}
\label{eq9}
\end{equation}
where $z$ is the three-body energy. To find the two-body energy, we should reduce it by spectator 
particle energy $E_{i}(\vec{p}_{i})$. Here, $\vec{k}_{i}$ is the relative momentum between the 
interacting pair ($jk$). The quantities $\mathcal{K}_{i,j}$ are Faddeev amplitudes, which are derived 
from Faddeev equation (\ref{eq4}). Since the $\pi{N}$ interaction is neglected, in this equation the 
Faddeev transition amplitudes corresponding to $\pi{N}$ system are missing. Using Eq.(\ref{eq9}), we 
define the transition probability of $(Y_{K})_{I=0}+N\rightarrow\pi\Sigma{N}$ as follows,
\begin{equation}
\begin{split}
& w(p'_{N},z)=\int d^{3}p_{N}\int d^{3}k_{N}\delta(z-Q(p_{N},k_{N})) \\
& \hspace{2cm}\times|T_{\pi\Sigma N\leftarrow(Y_{K})_{I=0}+N}(\vec{k}_N,\vec{p}_N,p'_{N};z)|^2.
\end{split}
\label{eq10}
\end{equation}
where $Q(p_{N},k_{N})$ is given by
\begin{equation}
Q(p_{N},k_{N})=\frac{p^{2}_{N}(m_{N}+m_{\pi}+m_{\Sigma})}{2m_{N}(m_{\pi}+m_{\Sigma})}-
\frac{k^{2}_{N}(m_{\pi}+m_{\Sigma})}{2m_{\pi}m_{\Sigma}}
\end{equation}
%%%%%%%%%%%%%%%%%%%%%%%%%%%%%%%%%%%%%%%%%%%%%%%%%%%%%%%%%%%%%%%%%%%%%%%%%%%%%%%
\section{TWO-BODY INPUT}
\label{put}
%%%%%%%%%%%%%%%%%%%%%%%%%%%%%%%%%%%%%%%%%%%%%%%%%%%%%%%%%%%%%%%%%%%%%%%%%%%%%%%
In this section, we give a brief survey on two-body interactions, which are the central inputs 
for the present few-body calculations. The $\bar{K}N-\pi\Sigma$ interaction is dominated by the 
$s$-wave $\Lambda$(1405) resonance. Therefore, the orbital angular momentum for $\bar{K}N$ 
interaction is taken to be zero. The $NN$ and $\Sigma{N}$ interactions were also taken in $l=0$ 
state. Since, the interaction in $\pi{N}$ subsystem is dominated by the $p$-wave component. 
Thus, in our few-body calculations the $\pi{N}$ interaction is neglected. All separable potentials 
in momentum representation have the form of Eq. (\ref{eq1}).
%%%%%%%%%%%%%%%%%%%%%%%%%%%%%%%%%%%%%%%%%%%%%%%%%
\subsection{$\bar{K}N-\pi\Sigma$ coupled-channel system}
\label{NN1}
%%%%%%%%%%%%%%%%%%%%%%%%%%%%%%%%%%%%%%%%%%%%%%%%%
During the past two decades, different phenomenological~\cite{dalit,akaishi,shev3} and chiral 
based~\cite{chi1,chi2,chi3,chi4,chi5,chi6,chi7} potentials are constructed to describe the 
$\bar{K}N$ interaction. The phenomenological models of interaction consider the $\Lambda(1405)$ 
resonance as a quasi-bound in $\bar{K}N$ system embedded in the $\pi\Sigma$ continuum. The 
chiral SU(3) dynamics has also turned out to be a successful approach to describe the $\bar{K}N$ 
interaction and the $\Lambda$(1405) resonance~\cite{chi1,chi2,chi3,chi4,chi5,chi6,chi7}. 
At and above $\bar{K}N$ threshold, the phenomenological and the chiral SU(3) $\bar{K}N$ 
models of interaction produce comparable results, while for subthreshold energies their 
results are different. The phenomenological $\bar{K}N$ interactions are constructed to 
reproduce a single pole nature for the $\Lambda$(1405) resonance as a quasi-bound state 
of the $\bar{K}N$ system around 1405 MeV. The $\bar{K}N-\pi\Sigma$ coupled-channels 
amplitude resulting from chiral SU(3) based potentials has two poles, one of them is 
located around 1420 MeV~\cite{chi2}. while the other pole with large width is located above 
the $\pi\Sigma$ threshold. Therefore, the chiral based potentials produce a binding 
energy of about 15 MeV for $\bar{K}N$ system, which is about half the binding produced 
with the purely phenomenological $\bar{K}N$ models of interaction.

We used different models to describe the s-wave $\bar{K}N-\pi\Sigma$ interaction, 
which is the most 
important interaction in the present three-body calculations with $\bar{K}NN$ and $\pi\Sigma{N}$ 
coupled-channels. We considered four types of phenomenological potentials. They reproduce the one- 
and two-pole structure for $\Lambda(1405)$ resonance and their parameters are given in 
Refs.~\cite{shev3,shev4}. The parameters are adjusted to reproduce all existing data on low-energy 
$\bar{K}N$ interaction. The $\bar{K}N-\pi\Sigma$ potentials in Ref.~\cite{shev4} are adjusted to 
reproduce the experimental results of the SIDDHARTA experiment~\cite{bazzi}. Depending on a pole 
structure of the $\Lambda(1405)$, we refer these potentials as \textquotedblleft{SIDD}, 
one-pole\textquotedblright and \textquotedblleft{SIDD}, two-pole\textquotedblright potential. The 
parameters of the potentials in Ref.~\cite{shev3} are adjusted to reproduce the experimental results 
of the KEK experiment~\cite{kek1,kek2}. Depending on a pole structure of the $\Lambda(1405)$, we 
refer these potentials as \textquotedblleft{KEK}, one-pole\textquotedblright and \textquotedblleft{KEK}, 
two-pole\textquotedblright potential.

Plus the phenomenological potentials, we also used two chiral based $\bar{K}N-\pi\Sigma$ 
potentials, which are given in Ref.~\cite{ohnishi}. 
These chiral based potentials reproduce the elastic and inelastic cross sections for the $K^{-}p$ 
reaction as well as the $\pi\Sigma$ mass spectra.
%%%%%%%%%%%%%%%%%%%%%%%%%%%%%
\subsection{$NN$ interaction}
\label{NN2}
%%%%%%%%%%%%%%%%%%%%%%%%%%%%%
In order to 
investigate the dependence of the $\pi\Sigma{N}$ invariant mass on nucleon-nucleon interaction models, 
we used two different potentials for $NN$ interaction. The first one is a two-term separable potential~\cite{gal}
\be
V_{NN}^{I}=\sum_{m=1}^{2}g_{NN}^{I,m}(k)\lambda_{NN}^{I,m}g_{NN}^{I,m}(k'),
\label{eq11}
\ee
where $\lambda_{NN}^{I,1}$ is negative  to take into account the short range repulsion part of the 
interaction. The parameters of this potential are given in Ref.~\cite{gal}.

We also used one-term PEST potential from Ref.~\cite{pest}. The strength parameter of the PEST 
potential is equal to one and the form-factor is defined by
\be
g_{NN}^{I}(k)=\frac{1}{2\sqrt{\pi}}\sum_{i=1}^{6}\frac{c_{i;I}^{NN}}{(\beta_{i;I}^{NN})^{2}+k^{2}},
\label{eq12}
\ee
where the parameters of the potential are given in Ref.~\cite{pest}. The PEST potential is not 
repulsive at short distances, but at low energies its phase shifts are close to the rank-two potential.

For the $s$-wave $\Sigma{N}$ interaction, we follow the form given in Ref.~\cite{tores},
\be
V_{\alpha\beta}^{I}(k^{\alpha},k^{\beta})=-\frac{C_{\alpha\beta}^{I}}{2\pi^2}(\Lambda_\alpha\Lambda_\beta)^{3/2}
(\mu_\alpha\mu_\beta)^{-1/2}g_{\alpha}^{I}(k^{\alpha})g_{\beta}^{I}(k^{\beta}).
\label{eq13}
\ee
where the form factor defined by $g_{\alpha}^{I}(k^{\alpha})=1/((k^{\alpha})^{2}+\Lambda_{\alpha}^{2})$. 
The coupling constants, $C_{\alpha\beta}^{I}$, and the range parameters $\Lambda_{\alpha}$ are given in 
Ref.~\cite{tores}.
%%%%%%%%%%%%%%%%%%%%%%%%%%%%%%%%%%%%%%%%%%%%%%%%%%%%%%%%%%%%%%%%%%%%%%%%%%%%%%%
\section{RESULTS AND DISCUSSION}
\label{result}
%%%%%%%%%%%%%%%%%%%%%%%%%%%%%%%%%%%%%%%%%%%%%%%%%%%%%%%%%%%%%%%%%%%%%%%%%%%%%%%
Solutions of the Faddeev equations corresponding to bound states and resonance poles in the 
$(I,J^{\pi})=(\frac{1}{2},0^{-})$ channel of the $\bar{K}NN-\pi\Sigma{N}$ three-body system 
were found by applying search procedures described in Sec.~\ref{formula}. In Table \ref{ta1} 
and \ref{ta2}, the results of the present work for three-body $\bar{K}NN-\pi\Sigma{N}$ quasi-bound 
state are presented. The sensitivity of the $\bar{K}NN$ pole position to the $\bar{K}N-\pi\Sigma$ 
interaction is investigated by using different potential models. In Table \ref{ta1} the pole 
position of the quasi-bound states in the $\bar{K}NN$ systems is calculated for phenomenological 
models of the $\bar{K}N-\pi\Sigma$ interaction and in Table \ref{ta2}, we calculated the pole 
energies for energy-dependent and energy-independent chiral potentials.
%%%%%%%%%%%%%%%%%%%%%%%%%%%%%%%%%%%%%%%%%%%%%%%%%%%%%%%%%%%%%%%%%%%%%
\begin{table}[H]
\caption{The sensitivity of the pole position(s) (in MeV), $z^{pole}_{X}$, of the $\bar{K}N$ 
and $\bar{K}NN$ systems to the different phenomenological models of the $\bar{K}N-\pi\Sigma$ 
interaction is investigated. $V^{One-pole}_{\bar{K}N-\pi\Sigma}$ and 
$V^{Two-pole}_{\bar{K}N-\pi\Sigma}$ standing for a one- and a two-pole structure of the 
$\Lambda$(1405) resonance, respectively.}
\centering
\begin{tabular}{ccc}
\hline\hline\noalign{\smallskip}\noalign{\smallskip}
 & $V^{One-pole}_{\bar{K}N-\pi\Sigma}$ & $V^{Two-pole}_{\bar{K}N-\pi\Sigma}$ \\
\noalign{\smallskip}\noalign{\smallskip}\hline
\noalign{\smallskip}\noalign{\smallskip}
SIDD pot.~\cite{shev3}: & & \\
\noalign{\smallskip}
$z^{pole}_{\bar{K}N}$  & $1428.1-i46.6$ & $1418.1-i56.9$ \\
\noalign{\smallskip}\noalign{\smallskip}
                       &                & $1382.0-i104.2$ \\
\noalign{\smallskip}\noalign{\smallskip}
$z^{pole}_{\bar{K}NN}$  & $2320.7-i31.5$ & $2325.0-i24.1$ \\
\noalign{\smallskip}\noalign{\smallskip}
\hline
\noalign{\smallskip}\noalign{\smallskip}
KEK pot.~\cite{shev4}: & & \\
\noalign{\smallskip}
$z^{pole}_{\bar{K}N}$ & $1411.3-i35.8$ & $1410.8-i35.9$ \\
\noalign{\smallskip}\noalign{\smallskip}
                      &                & $1380.8-i104.8$ \\
\noalign{\smallskip}\noalign{\smallskip}
$z^{pole}_{\bar{K}NN}$ & $2329.0-i26.0$ & $2327.6-i19.5$ \\
\noalign{\smallskip}\noalign{\smallskip}
\hline\hline
\end{tabular}
\label{ta1} 
\end{table}
%%%%%%%%%%%%%%%%%%%%%%%%%%%%%%%%%%%%%%%%%%%%%%%%%%%%%%%%%%%%%%%%%%%%%
%%%%%%%%%%%%%%%%%%%%%%%%%%%%%%%%%%%%%%%%%%%%%%%%%%%%%%%%%%%%%%%%%%%%%
\begin{table}[H]
\caption{The pole position(s) (in MeV), $z^{pole}_{X}$, of the quasi-bound states in the $\bar{K}N$ 
and $\bar{K}NN$ systems is calculated for energy-dependent and energy-independent chiral potentials.}
\centering
\begin{tabular}{ccc}
\hline\hline\noalign{\smallskip}\noalign{\smallskip}
 & $V^{E-indept.}_{\bar{K}N-\pi\Sigma}$ & $V^{E-dept.}_{\bar{K}N-\pi\Sigma}$ \\
\noalign{\smallskip}\noalign{\smallskip}\hline
\noalign{\smallskip}\noalign{\smallskip}
%chiral potential~\cite{}: & & \\
\noalign{\smallskip}
$z^{pole}_{\bar{K}N}$ \, \,  & $1407.2-i18.5$ \, \, & $1420.6-i20.3$ \\
\noalign{\smallskip}\noalign{\smallskip}
                      \, \,  &                \, \, & $1343.0-i72.5$ \\
\noalign{\smallskip}\noalign{\smallskip}
$z^{pole}_{\bar{K}NN}$ \, \,  & $2313.4-i21.9$ \, \, & $2346.5-i22.0$ \\
\noalign{\smallskip}\noalign{\smallskip}
\hline\hline
\end{tabular}
\label{ta2} 
\end{table}
%%%%%%%%%%%%%%%%%%%%%%%%%%%%%%%%%%%%%%%%%%%%%%%%%%%%%%%%%%%%%%%%%%%%%

The position of a quasi-bound state in the three-body problem is usually defined by solving the 
homogeneous integral equations (\ref{eq8}) which comes from the separable expansion of the Faddeev 
amplitudes. To find the resonance energy of the system using these equations, one should search 
for a complex energy at which one of the eigenvalues of the kernel matrix becomes equal to one. 
Therefore, as one can see from Eqs.~\ref{eq6} and \ref{eq7}, the Faddeev amplitudes will have a 
pole at this energy.

To examine the efficiency of the separable expansion method, we used another way to find the 
$K^{-}pp$ pole position(s) without using the integration in the complex momentum plane. The 
signal of the $K^{-}pp$ bound state would be seen in the Faddeev transition amplitudes. In the 
present work, we studied how the signature of the $K^{-}pp$ quasi-bound state shows up in the 
three-body scattering amplitudes by using coupled-channel Faddeev AGS equations. To achieve 
this goal, we must solve the inhomogeneous integral equations for the amplitudes defined in 
Eq.~(\ref{eq4}).

\begin{figure}[H]
\begin{center}
\includegraphics[scale=0.6]{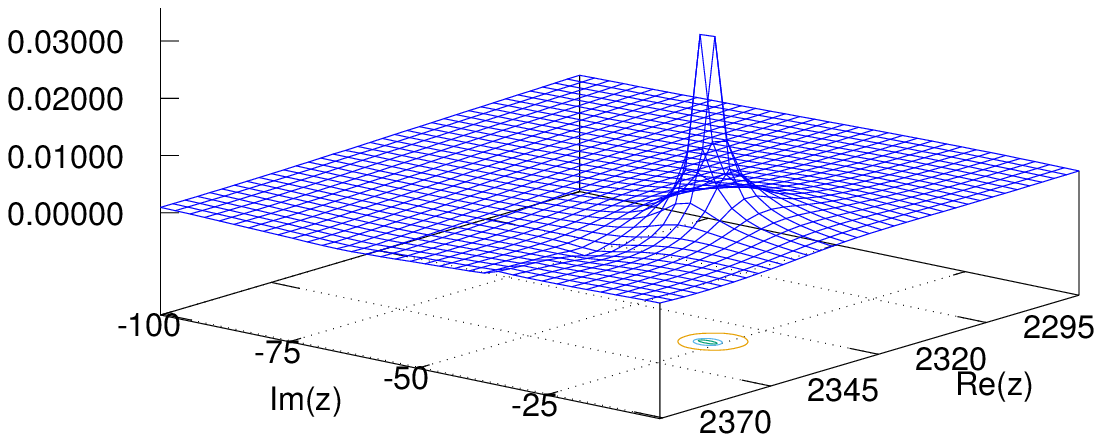}
\includegraphics[scale=0.6]{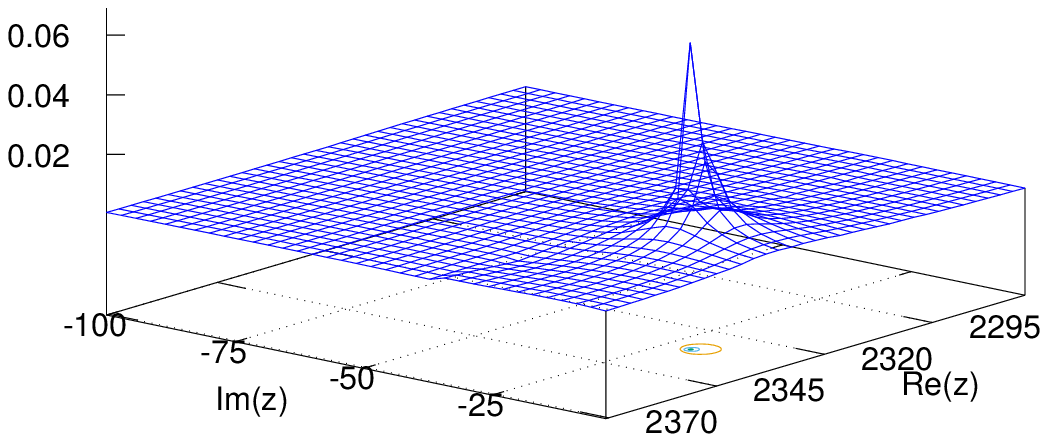} \\
\includegraphics[scale=0.6]{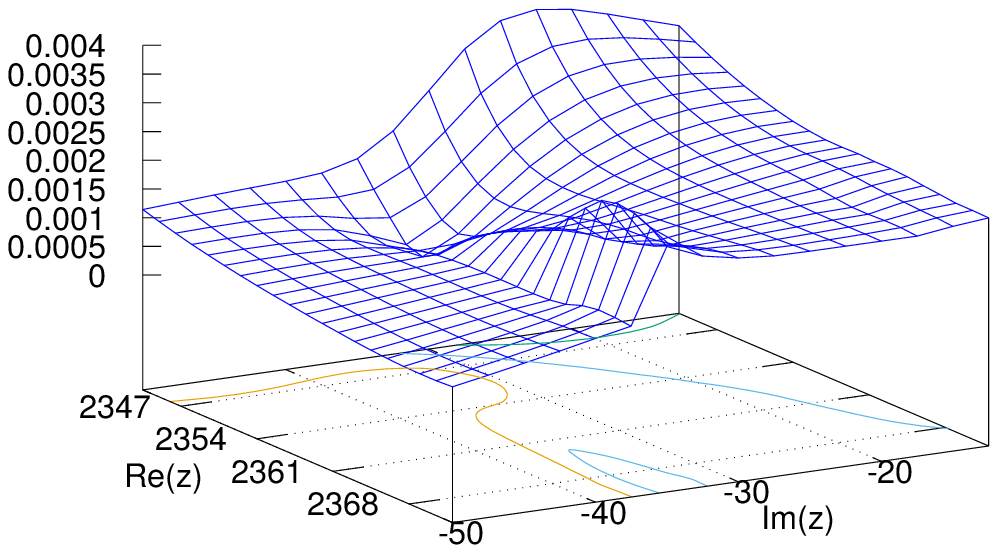}
\includegraphics[scale=0.6]{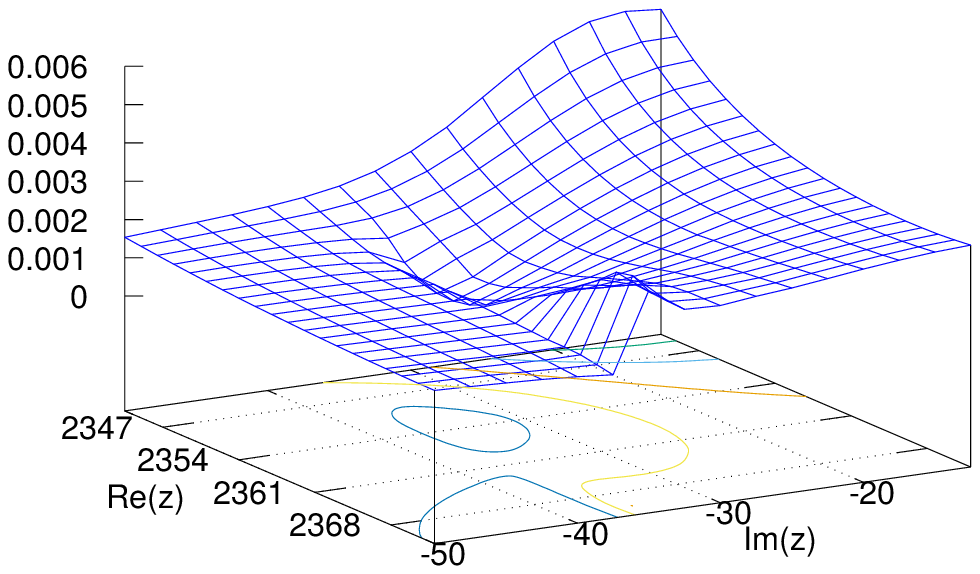} \\

\end{center}
\vspace{-0.5cm}
\caption{(Color online) The Faddeev amplitude for $(\bar{K}N)_{I=0}+N\rightarrow (\bar{K}N)_{I=0}+N$ 
reaction. The transition amplitudes calculated using one-pole (left) and two-pole (right) models 
of KEK potential. In second row, the imaginary part of $z$, was chosen to be between -50 MeV and 
-12 MeV and the real part change from 2347 MeV to 2375 MeV to make the branch points more visible. 
The three-body calculations performed by using the PEST potential model for $NN$ interaction.}
\label{three-kek}
\end{figure}

Fig.~\ref{three-kek} shows the calculated three-body scattering amplitude 
$|\mathcal{K}_{(\bar{K}N)_{I=0}+N\rightarrow (\bar{K}N)_{I=0}+N}(p_{N},p'_{N};z)|$ whose initial and 
final states are $(\bar{K}N)_{I=0}+N$. The off-shell momenta $p$ and $p'$ are equal, 150 MeV/c and 
the real and imaginary part of the three-body energy, $z$, change from -100 MeV to 0 MeV. We used 
one-pole (left) and two-pole (right) version of the KEK potential for describing the $\bar{K}N-\pi\Sigma$ 
interaction. Since the input energy of the AGS equations is complex the moving singularities which are 
caused by the open channel $\pi\Sigma{N}$, will not appear in the three-body amplitudes. The calculated 
resonance energies of the $\bar{K}NN$ system by this method, have presented in Table~\ref{ta-amp}. 
Comparing these results with those in Table~\ref{ta1}, one can see that both results are in good 
agreement with each other.

When at least one of the intermediate particles is unstable, plus the signal of the resonance states, 
one can see a branch point in the complex plane. In Fig.~\ref{three-kek}, plus the signature of $K^{-}pp$ 
pole position, we can see the branch points i.e., a threshold opening associated with the $\Lambda(1405)$ 
pole, situated at $z=M_{N}+M_{\Lambda(1405)}$, sum of nucleon mass and $\Lambda(1405)$ pole position. 
In the second row of Fig.~\ref{three-kek}, we have shown the branch points for $\Lambda(1405)N$ intermediate 
state. The branch point is clearly visible, together with the cut that in this picture is chosen in the 
positive Re $z$ direction. In the second row, to make the branch points more visible the imaginary part of 
the three-body energy, $z$, was chosen to be between -50 MeV and -12 MeV and the real part change from 2347 
MeV to 2375 MeV. We used one-pole (left) and two-pole (right) version of the KEK potential to extract the 
branch points. 
%%%%%%%%%%%%%%%%%%%%%%%%%%%%%fig.2
%%%%%%%%%%%%%%%%%%%%%%%%%%%%
%%%%%%%%%%%%%%%%%%%%%%%%%%%%%
\begin{table}[H]
\caption{The pole position (in MeV) of the Faddeev amplitude for 
$(\bar{K}N)_{I=0}+N\rightarrow (\bar{K}N)_{I=0}+N$ reaction. The pole positions are calculated 
for one- and two-pole models of KEK potential.}
\centering
\begin{tabular}{ccc}
\hline\hline\noalign{\smallskip}\noalign{\smallskip}
 & \,\,\,\, $V^{KEK, One-pole}_{\bar{K}N-\pi\Sigma}$ \,\,\,\, & \,\,\,\, $V^{KEK, Two-pole}_{\bar{K}N-\pi\Sigma}$ \\
\noalign{\smallskip}\noalign{\smallskip}\hline
\noalign{\smallskip}\noalign{\smallskip}
$z^{pole}_{\bar{K}NN}$ \,\,\,\, & \,\,\,\, $ 2331.0-i27.2$ \,\,\,\, & \,\,\,\, $2328.5-i19.8$ \\
\noalign{\smallskip}\noalign{\smallskip}
\hline\hline
\end{tabular}
\label{ta-amp} 
\end{table}
%%%%%%%%%%%%%%%%%%%%%%%%%%%%%

We investigated the dependence of the two- and three-body pole energy trajectories on the magnitude 
$\lambda^{I=0}_{\bar{K}N-\bar{K}N}$, when the $\bar{K}N$ strength parameter is increased from its 
physical value. Let $\kappa$ stand for an enhancement factor of strength of the $I=0$ $\bar{K}N$ interaction:
\be
\bar{\lambda}_{\bar{K}N-\bar{K}N}^{I=0}=\kappa \lambda_{\bar{K}N-\bar{K}N}^{I=0}.
\label{eq14}
\ee

\begin{figure}[H]
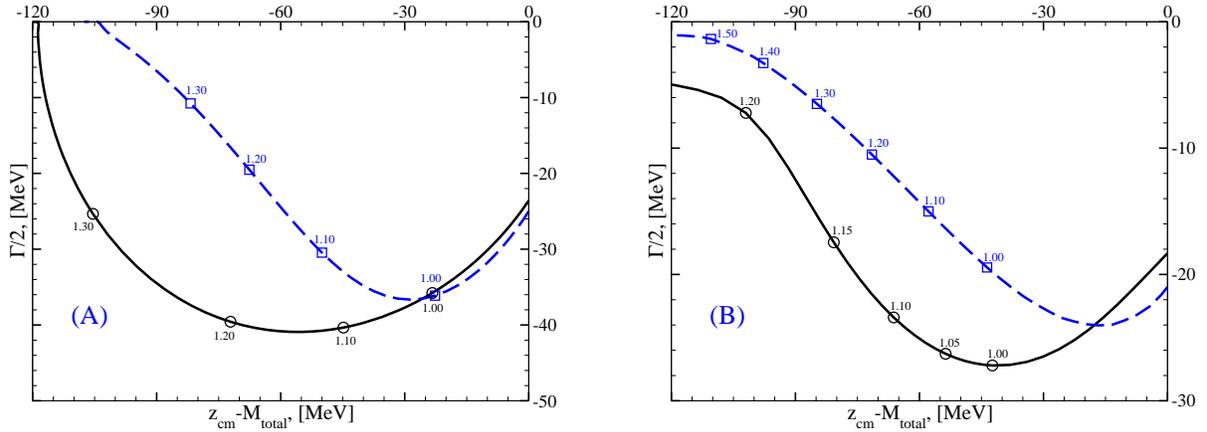

\begin{center}
\includegraphics[scale=0.35]{kp.eps}
\hspace{1cm}
\includegraphics[scale=0.35]{kpp.eps}
\end{center}
\vspace{-0.5cm}
\caption{(Color online) Resonance energy of $\bar{K}N$ (A) and $\bar{K}NN-\pi\Sigma{N}$ 
(B) systems. The numbers attached to the squares and circles give the corresponding values of the enhancement 
factor $\kappa$ in Eq. (\ref{eq14}). The resonance energies calculated using KEK potentials for 
$\bar{K}N-\pi\Sigma$ interaction given in Ref.~\cite{shev3}. These potential models can reproduce the one- and 
two-pole structure of $\Lambda$(1405) resonance. The solid lines show the resonance energies with one-pole model, 
and the dashed lines show the resonance energies with the two-pole model of the KEK potential. Here, $z_{cm}$ is 
the energy of the system in center of mass frame and $M_{total}$ is the mass of the two- and three-body system.}
\label{kpp-pole-tra}
\end{figure}

We calculated the pole trajectory for one- and two-pole version of the KEK potential. The behavior of the 
two-body $K^{-}p$ and three-body $K^{-}pp$ pole energy trajectories are quite different at $\pi\Sigma$ 
threshold. The pole energies obtained for three-body system are shown in Fig. \ref{kpp-pole-tra} (B). 
The blue dashed and black solid curves in Fig.~\ref{kpp-pole-tra} (A) correspond to the two-body 
results. The numbers attached to the circles and squares give the corresponding values of the enhancement 
factor $\kappa$. As $\kappa$ increases, the binding energy of the system increases for both the two- and 
three-body systems. In the two-body calculations of two-pole potential, the imaginary part of the resonance 
energy becomes smaller as the binding energy increases and for $\kappa\sim$1.33 (at $\pi\Sigma$ threshold) 
the resonance almost becomes a bound state in $\pi\Sigma$ channel. In contrary, in the three-body system 
the resonance energy will have a non zero imaginary part at the $\pi\Sigma$ threshold as $\kappa$ grows, 
since the $\pi\Lambda$ channel is included effectively.
%%%%%%%%%%%%%%%%%%%%%%%%%%%%%fig.3
%%%%%%%%%%%%%%%%%%%%%%%%%%%%
%%%%%%%%%%%%%%%%%%%%%%%%%%%%%
%%%%%%%%%%%%%%%%%%%%%%%%%%%%%
\subsection{Calculation of the transition probability}
\label{tranpo1}
\begin{figure}[H]
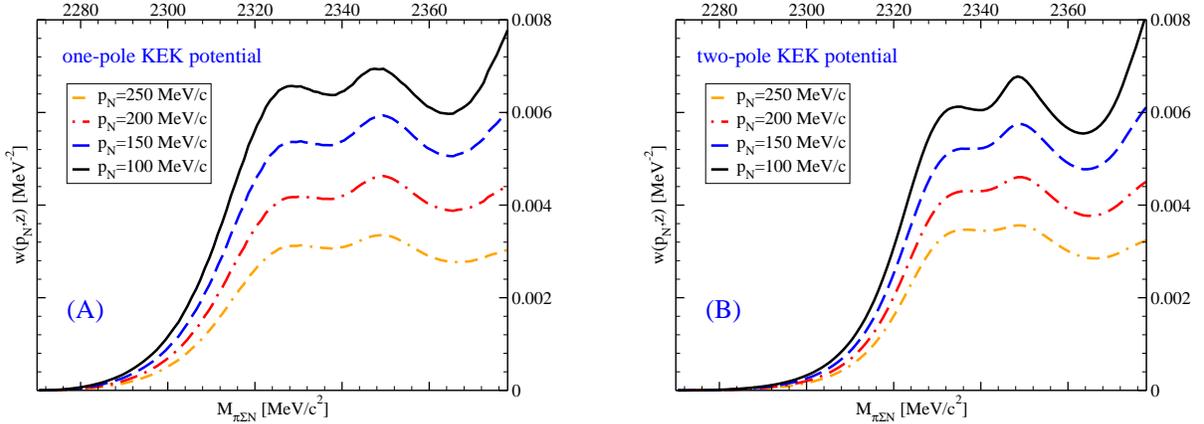

\begin{center}
\includegraphics[scale=0.35]{kek-onepole.eps}
\hspace{1cm}
\includegraphics[scale=0.35]{kek-twopole.eps}
\end{center}
\vspace{-0.5cm}
\caption{(Color online) The $\pi\Sigma{N}$ mass spectra for 
$(Y_{K})_{I=0}+N\rightarrow\pi\Sigma{N}$ reaction. The $M_{\pi\Sigma{N}}$ spectra calculated using 
KEK potentials for $\bar{K}N-\pi\Sigma$ interaction given in Ref.~\cite{shev3}. The calculations 
are performed by using the PEST potential for $NN$ interaction. To investigate the energy dependence 
of the transition probability, we calculated $w(p_{N},z)$ for $p_{N}=100-250$ MeV/c.}
\label{shev-pest1}
\end{figure}

\begin{figure}[H]
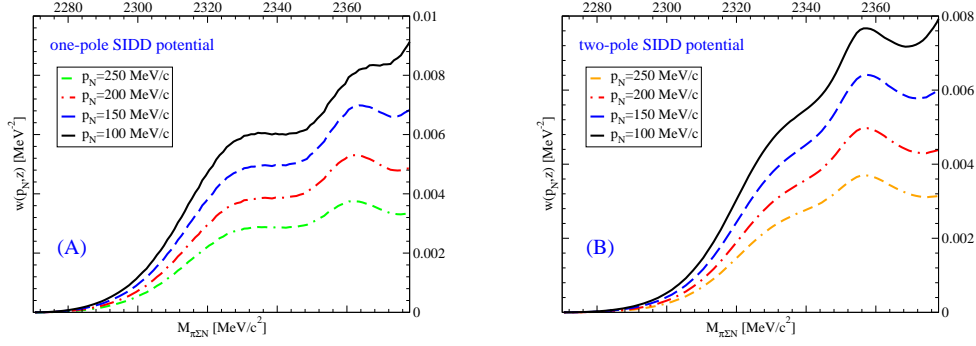

\begin{center}
\includegraphics[scale=0.28]{sidd-onepole.eps}
\hspace{1cm}
\includegraphics[scale=0.28]{sidd-twopole.eps}
\end{center}
\vspace{-0.5cm}
\caption{(Color online) Same as Fig.\ref{shev-pest1} but using SIDDHARTA potentials~\cite{shev4} 
for $\bar{K}N-\pi\Sigma$ interaction which reproduce the one- and two-pole structure of 
$\Lambda$(1405) resonance.}
\label{shev-pest2}
\end{figure}

\begin{figure}[H]
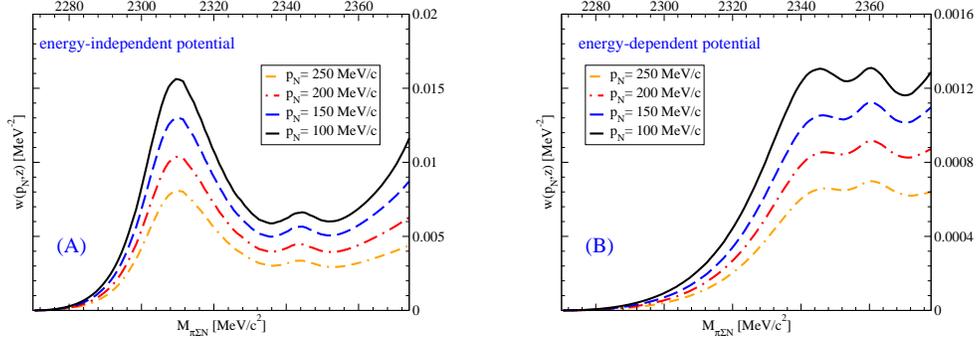

\centering
\includegraphics[scale=0.28]{in-5mev.eps}
\hspace{1cm}
\includegraphics[scale=0.28]{de-5mev.eps}
\caption{(Color online) Same as Fig.\ref{shev-pest1} but for the energy-dependent (B) and 
energy-independent (A) potentials given in Ref.~\cite{ohnishi}.}
\label{ohni-pest1}
\end{figure}

The calculated resonance energies that have presented in Table \ref{ta1} and \ref{ta2}, give only pole 
positions of the $K^{-}pp$ system. However, we know that these results are not a quantity that can be 
directly measured in any experiments. To examine the existence of the quasi-bound state in $K^{-}pp$ 
system by experiments, one has to calculate the cross sections of $K^{-}pp$ production reactions. We 
can use the calculated results in Table~\ref{ta1} and~\ref{ta2} and also in Fig.~\ref{kpp-pole-tra} as 
guideline to study these reactions. As it was said in Sect.~\ref{intro}, the $K^{-}pp$ quasi-bound 
state can be produced through kaon-induced reactions on light nuclei such as $\mathrm{^{3}{He}}$ 
and deuteron. The trace of the resonances would be seen in the mass spectra of the final particles. 
In the present calculations, we studied how good the signature of the $K^{-}pp$ system shows up in the 
observables of the three-body reactions by using coupled-channel Faddeev equations in the AGS form. To 
achieve this goal, we must solve the coupled integral equations for the amplitudes defined in Eq.~(\ref{eq4}), 
and then construct the breakup amplitudes $T_{\pi\Sigma{N}\leftarrow(Y_{K})_{I=0}+N}$ defined in Eq.~(\ref{eq9}). 
Since the kernel of AGS equations has the standard moving singularities that are caused by the opened 
channel $\pi\Sigma{N}$ and are encountered in any three-body breakup problem, we have followed the same 
procedure implemented in Refs.~\cite{poin1,poin2}. Using the so called \textquotedblleft point-method
\textquotedblright, we computed the cross section of $(Y_{K})_{I=0}+N\rightarrow\pi\Sigma{N}$ reaction 
and studied the behavior of $\pi\Sigma{N}$ mass spectra. The transition probabilities for phenomenological 
potentials are depicted in Figs. \ref{shev-pest1} and \ref{shev-pest2}. In Fig.~\ref{ohni-pest1} the 
three-body calculations are performed by chiral based potentials for $\bar{K}N$ interaction and PEST potential 
for $NN$ interaction.
%%%%%%%%%%%%%%%%%%%%%%%%%%%%%fig.4
%%%%%%%%%%%%%%%%%%%%%%%%%%%%%
%%%%%%%%%%%%%%%%%%%%%%%%%%%%%fig.5
%%%%%%%%%%%%%%%%%%%%%%%%%%%%
%%%%%%%%%%%%%%%%%%%%%%%%%%%%%fig.6
%%%%%%%%%%%%%%%%%%%%%%%%%%%%%%%%%%%%%%%
%%%%%%%%%%%%%%%%%%%%%%%%%%%%%%%%%%%%%%%%%%%%%%%%%%%%%%%%%%
\begin{table}[H]
\caption{The dependence of the $\bar{K}N$ (first pole) and $\bar{K}NN$ quasi-bound state positions 
(in MeV) on the $\lambda_{\bar{K}N-\bar{K}N}$ strength parameter in the $I=0$ state. We calculated 
the pole positions for three values of $\kappa$ parameter.}
\centering
\begin{tabular}{cccc}
\hline\hline\noalign{\smallskip}\noalign{\smallskip}
 & $\kappa=1.00$ & $\kappa=1.05$ & $\kappa=1.10$ \\
\noalign{\smallskip}\noalign{\smallskip}\hline
\noalign{\smallskip}\noalign{\smallskip}
\noalign{\smallskip}
$z^{pole}_{\bar{K}N}$ \,   & $1420.6-i20.25$ \, & $1414.4-i21.8$ \, & $1407.7-i24.0$ \\
\noalign{\smallskip}\noalign{\smallskip}
$z^{pole}_{\bar{K}N(+N)}$  \,  & $2359.5-i20.25$ \, & $2353.3-i21.8$ \, & $2346.6-i24.0$ \\
\noalign{\smallskip}\noalign{\smallskip}
$z^{pole}_{\bar{K}NN}$  \,  & $2346.5-i22.0$ \,  & $2339.0-i22.1$ \,  & $2331.6-i21.5$ \\
\noalign{\smallskip}\noalign{\smallskip}
\hline\hline
\end{tabular}
\label{ta3} 
\end{table}
%%%%%%%%%%%%%%%%%%%%%%%%%%%%%%%%%%%%%%%%%%%%%%%%%%%%%%%%%%

In Fig. \ref{shev-pest1}, we calculated the $\pi\Sigma{N}$ mass spectra using one-pole (A) and 
two-pole (B) version of KEK potentials for $\bar{K}N-\pi\Sigma$ interaction given in Ref.~\cite{shev3}. 
To investigate the energy dependence of the transition probability, we calculated $w(p_{N},z)$ for 
$p_{N}=100-250$ MeV/c. We investigated the dependence of $\pi\Sigma{N}$ mass spectra on two-body 
$\bar{K}N-\pi\Sigma$ interactions, necessary for the description of the $\bar{K}NN-\pi\Sigma{N}$ 
system. Therefore, in Fig. \ref{shev-pest2}, we calculated the $\pi\Sigma{N}$ mass spectra using 
the one- and two-pole version of the SIDDHARTA potential for $\bar{K}N-\pi\Sigma$ interaction given 
in Ref.~\cite{shev3}. The results suggest that a distinct peak of bound kaonic states should be observed, 
regardless of the momentum value and the class of the $\bar{K}N-\pi\Sigma$ interaction. In the calculated 
mass spectra for the two-pole model of the KEK and SIDDHARTA potentials, the second pole of $\Lambda(1405)$ 
resonance with its large width, does not affect the $\pi\Sigma{N}$ invariant mass. As one can see from 
Figs.~\ref{shev-pest1} and~\ref{shev-pest2}, all potential models will produce the mass spectra with the 
similar behavior and two distinct bump structures can be seen in the $\pi\Sigma{N}$ invariant mass.

The results of the full coupled-channel calculations of the $(Y_{K})_{I=0}+N\rightarrow\pi\Sigma{N}$ 
scattering using two versions of the energy-dependent and energy-independent $\bar{K}N-\pi\Sigma$ 
potentials derived based on chiral SU(3) dynamic and non relativistic kinematics are shown in 
Fig.~\ref{ohni-pest1}. It is seen, that the three-body results corresponding to each version of $\bar{K}N$ 
interaction differ sufficiently. Therefore, in principle, it would be possible to favor one version of 
the $\bar{K}N-\pi\Sigma$ potential by comparing with experimental results.
Within this model, we have found two bump structures appearing in the $(Y_{K})_{I=0}+N\rightarrow\pi\Sigma{N}$ 
transition probabilities in the energy region around the $\bar{K}NN$ pole position and 
$z=M_{N}+M_{\Lambda(1405)}$. As it was said before, the second bump which is situated at $z=M_{N}+M_{\Lambda(1405)}$ 
is actually originated from a branch point in the complex plane (see Fig.\ref{three-kek}), i.e., a threshold 
opening associated with the $\Lambda(1405)$ pole.

\begin{figure}[H]
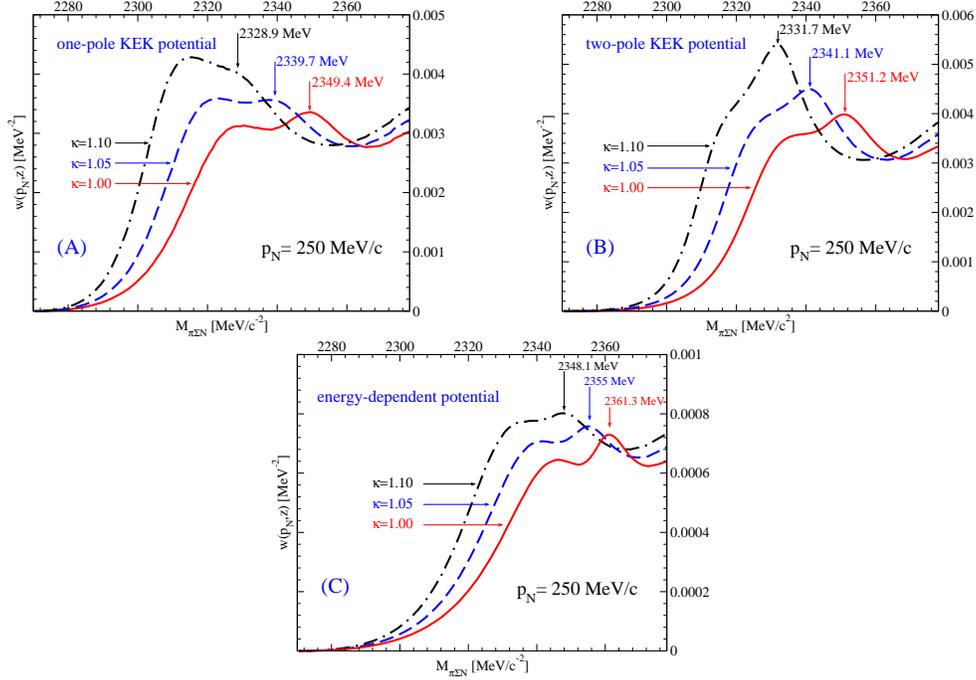

\begin{center}
\includegraphics[scale=0.28]{kek-onepole-tra.eps}
\hspace{1cm}
\includegraphics[scale=0.28]{kek-twopole-tra.eps} \\
\includegraphics[scale=0.28]{ohni-poletra.eps}
\caption{(Color online) Our results for $\pi\Sigma{N}$ mass spectra are presented using 
\textquotedblleft{KEK}, one-pole\textquotedblright, \textquotedblleft{KEK}, two-pole\textquotedblright 
and also energy-dependent potentials for $\bar{K}N$ interaction and PEST potential for $NN$ interaction. 
We calculated the $\pi\Sigma{N}$ invariant mass for three different values of the $\kappa$ coefficient. 
As one can see, when we increase the $\kappa$ parameter, the movement of the bumps location are very 
similar to the movement of the pole positions in the $\bar{K}N(+N)$ and $\bar{K}NN$ systems, which 
are presented in Table~\ref{ta3} and Fig.~\ref{kpp-pole-tra}.}
\end{center}
\label{shev-pest3}
\end{figure}

To show that, these bumps are really corresponding to the quasi-bound state in the $\bar{K}NN$ 
system and $\Lambda(1405)$ pole and are not caused by threshold effects. Let us investigate these 
bump structures in the $\pi\Sigma{N}$ mass spectra and clarify the origin of these bumps. In 
Fig. \ref{shev-pest3}, we calculated the $\pi\Sigma{N}$ mass spectra using one- and two-pole 
version of the KEK and also energy-dependent chiral potentials for $\bar{K}N-\pi\Sigma$ interaction 
when the magnitude of the $(I=0)$ $\bar{K}N$ strength parameter is increased from its physical 
value. We calculated the $\pi\Sigma{N}$ mass spectra for three values of the enhancement factor 
$\kappa=$1.0, 1.05 and 1.10. As it was shown in Fig.~\ref{kpp-pole-tra} and Table~\ref{ta3}, when 
we increase the $\kappa$ parameter, the binding energy of the system will increase for both the 
two- and three-body systems and the pole energies will go toward the $\pi\Sigma(+N)$ and $\pi\Sigma{N}$ 
threshold, respectively. Comparing the results of the mass spectra with those presented in 
Table~\ref{ta3} and Fig.~\ref{kpp-pole-tra}, one can see that the bump structures in the mass 
spectra and the quasi-bound states in Table~\ref{ta3} will locate at the same energy and have the 
same movement. Therefore, one can say that the first bump structure should be corresponding to a 
quasi-bound state in $\bar{K}NN$ system and the second bump structure is derived from a branch point 
in the complex plane.
%%%%%%%%%%%%%%%%%%%%%%%%%%%%%fig.7
%%%%%%%%%%%%%%%%%%%%%%%%%%%%%

\begin{figure}[H]
\centering
\includegraphics[scale=0.32]{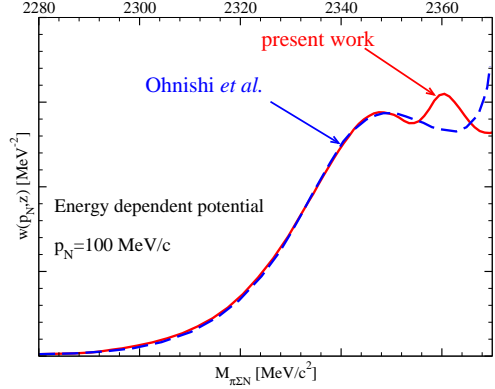}
\caption{(Color online) The calculated $\pi\Sigma{N}$ mass spectra for energy-dependent potential 
is compared with other theoretical results. The $\pi\Sigma{N}$ mass spectra for incident momentum 
$p_{N}=100$ MeV/c was calculated. Our result is shown by red solid curve and the result by Ohnishi 
{\it et al.}~\cite{ohnishi} by blue dashed curve.}
\label{ohni-gal1}
\end{figure}

In order to compare the present results with those in Ref.~\cite{ohnishi}, we calculated $w(p_{N},z)$ 
for $p_{N}=100$ MeV/c using the same $\bar{K}N-\pi\Sigma$ and a two-term type potential~\cite{gal} 
with a repulsive core and an intermediate-range attraction is used to describe the nucleon-nucleon 
interaction. The $\pi\Sigma{N}$ invariant mass obtained with the two-terms $V^{I}_{NN}$ are shown 
in Fig.~\ref{ohni-gal1}. Energy-dependent set of $\bar{K}N-\pi\Sigma$ potential was used together.

In contrast to the results of Ref.~\cite{ohnishi}, our results show that, plus the bump 
related to the quasi-bound state in $\bar{K}NN$ systems, a typical bump structure manifests 
itself in the $\pi\Sigma{N}$ invariant mass at the energy related to the quasi-bound state 
in $\bar{K}N(+N)$ system. However, this bump structure in the observables dose not derives 
from a resonance pole and the origin of this structure is the branch points. The behavior of 
the mass spectra is similar to the extracted results for the phenomenological potentials. 
The difference between the present results and those by Ohnishi {\it et al.,} can be important. 
In our results, we have two bump structure close to each other in the mass spectra, one is 
related to the quasi-bound state in $K^{-}pp$ system and the other corresponding to branch 
points which originates from intermediate $\Lambda(1405)$. Thus, this effect should be taken 
into account in theoretical interpretation of the experimental results by E15.
%%%%%%%%%%%%%%%%%%%%%%%%%%%%%fig.8
%%%%%%%%%%%%%%%%%%%%%%%%%%%%%
%%%%%%%%%%%%%%%%%%%%%%%%%%%%%%%%%%%%%%%%%%%%%%%%%
\subsection{Averaged transition probability}
\label{tranpo2}
In Subsection~\ref{tranpo1}, we calculated the transition probabilities for four discrete values of momentum, 
but, in actual situation the momentum $p'_{N}$ can occupy any value over a continuous range. To include all 
these momenta into consideration, in this subsection, we calculated the averaged transition probability 
$\bar{w}$, which is given by

\begin{equation}
\begin{split}
& \bar{w}(z)=\int d^{3}p_{N}\int d^{3}k_{N}\delta(z-Q(p_{N},k_{N})) \\
& \hspace{0.8cm}\times |\int p'^{2}_{N}dp'_{N} \rho(p'_{N}) T_{\pi\Sigma N
\leftarrow(Y_{K})_{I=0}+N}(\vec{k}_N,\vec{p}_N,p'_{N};z)|^2.
\end{split}
\label{eq15}
\end{equation}
where the function $\rho(p_{N})$ can be defined by
\begin{equation}
1=\langle\psi|\psi\rangle=\int_{0}^{\infty}p_{N}^{2}dp_{N}\rho(p_{N}).
\label{eq16}
\end{equation}

\begin{figure}[H]
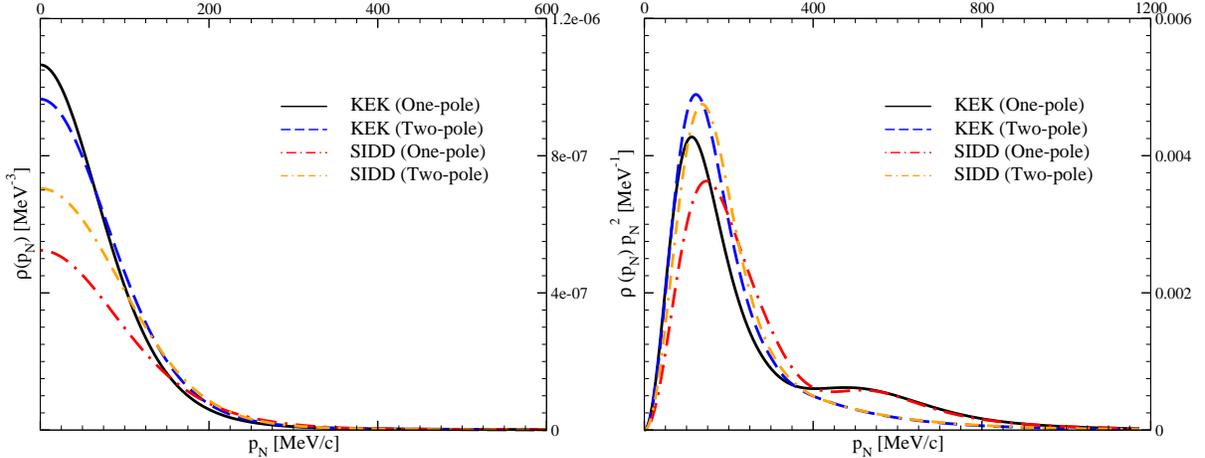

\begin{center}
\includegraphics[scale=0.38]{say2.eps}
\includegraphics[scale=0.38]{say2p2.eps}
\caption{(Color online) The momentum distribution of the spectator nucleon, (up) 
$\rho(p_{N})$ and (down) $p^{2}_{N}\rho(p_{N})$, in $K^{-}pp$ system. We used different kind of 
potentials to study the model dependence of the momentum distributions.}
\end{center}
\label{say1}
\end{figure}

To define the wave function of the $K^{-}pp$ three-body system, we used the so called "exact optical" 
potential~\cite{shev3}. Therefore the $\pi\Sigma{N}$ channel has not included directly into account 
and we can drop the particle channel indices. Using Faddeev equations in \ref{eq8}, we can calculate 
three Faddeev components, $\phi^{\alpha}_{i,I_{i}}(p_{i},k_{i};z)$, which are given by  
\begin{equation}
\begin{split}
& \phi_{i,I_{i}}(p_{i},k_{i};z)=\frac{1}{z-k_{i}^{2}/2\mu_{i}-p_{i}^{2}/2\nu_{i}} \\
& \hspace{1.7cm}\times g_{i,I_{i}}(k_{i})\tau_{i,I_{i}}(z-E_{i}(p_{i}))u_{n;i,I_i}(p_{i},z),
\end{split}
\label{eq17}
\end{equation}
where $\mu_{i}=\frac{m_{j}m_{k}}{m_{j}+m_{k}}$ is the reduced mass of the interacting pairs $jk$ and 
also $p_{i}$ and $k_{i}$ are the Jacobi momenta of the spectator particle and interacting particles, 
respectively. The $K^{-}pp$ three-body wave function can be defined by
\begin{equation}
|\psi\rangle=\sum_{i,I_{i}}|\phi_{i,I_{i}}\rangle,
\label{eq18}
\end{equation}
where $\psi$ is the normalized wave function of the $K^{-}pp$ system, which is defined as a sum 
of the above components. Figure~\ref{say1} (up) shows the momentum distribution of the spectator 
nucleon, $\rho_{N}(p)$, in $K^{-}pp$ system for various models of the $\bar{K}N-\pi\Sigma$ potential 
and Figure~\ref{say1} (down) shows the same distributions but multiplied by $p_{N}^{2}$.
%%%%%%%%%%%%%%%%%%%%%%%%%%%%%fig.9
[H]
%%%%%%%%%%%%%%%%%%%%%%%%%%%%%%%%%%%%%%%

\begin{figure}[H]
\centering
\includegraphics[scale=0.38]{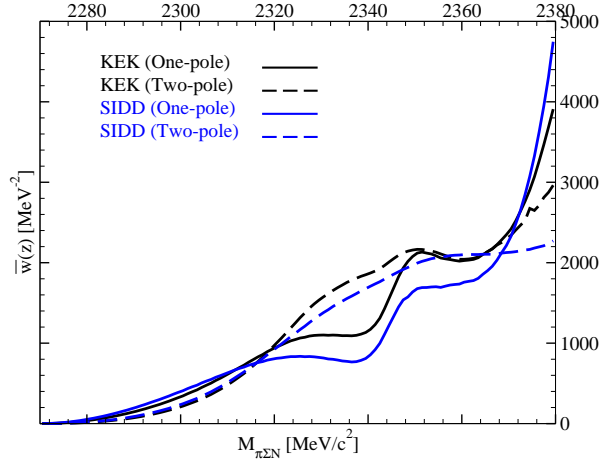}
\caption{(Color online) Energy dependence of the averaged transition probability, $\bar{w}(z)$. 
To study the model dependence of the results, we used KEK potential (black curves) and SIDD 
potentials (blue curves) in our calculations. The solid curves represent the results of the 
one-pole models and the dashed curves belong to the two-pole models.}
\label{say2}
\end{figure}

Fig.~\ref{say2} shows the calculated $\pi\Sigma{N}$ mass spectra using one- and two-pole version of 
KEK (black curves) and SIDD potential (blue curves) for $\bar{K}N-\pi\Sigma$ interaction. As one can 
see, the mass spectra around the $\bar{K}NN$ threshold are affected by the kinematical effects and the 
peaks corresponding to the $\Lambda(1405)+N$ branch point and $K^{-}pp$ quasi-bound 
states are not as clear as in Figs.~\ref{shev-pest1} and \ref{shev-pest2}. According to the 
Figs.~\ref{shev-pest1}, \ref{shev-pest2} and \ref{say1}, these changes were expected, because the 
threshold effects are stronger for low values of the $p'_{N}$ momentum and the weight of them in the 
momentum distribution are larger than high momentum.
%%%%%%%%%%%%%%%%%%%%%%%%%%%%%fig.10
%%%%%%%%%%%%%%%%%%%%%%%%%%%%%%%%%%%%%%%

In general, Faddeev equations need as input a potential that describes the interaction between 
two individual particles. It is also possible to introduce a term in the equation in order to 
take three-body forces into account. Although, we think, that while our information about 
the two-body $\bar{K}N$ interaction is not completed, the inclusion of three-body forces can not 
be necessary. One can also investigate the dependence of the mass spectra on the two-body local 
potentials~\cite{chi7}. The three-body theory of reactions can also be formulated for local potentials on 
the basis of the Faddeev equations. The work in this direction is underway and will be reported 
elsewhere.

%%%%%%%%%%%%%%%%%%%%%%%%%%%%%%%%%%%%%%%%%%%%%%%%%%%%%%%%%%%%%%%%%%%%%%%%%%%%%%%
\section{CONCLUSION}
\label{conc}
%%%%%%%%%%%%%%%%%%%%%%%%%%%%%%%%%%%%%%%%%%%%%%%%%%%%%%%%%%%%%%%%%%%%%%%%%%%%%%%
In summary, in this work exact Faddeev-type calculations of $\bar{K}NN-\pi\Sigma{N}$ system were 
performed to define the binding energy and width of $K^{-}pp$ system. The efficiency of the so 
called HSE method was investigated. We have calculated the transition probability (\ref{eq10}) for 
$(Y_{K})_{I=0}+N\rightarrow\pi\Sigma{N}$ reaction in the energy region between the $\bar{K}NN$ 
and $\pi\Sigma{N}$ thresholds. We have examined how the signature of the $K^{-}pp$ quasi-bound 
state in the three-body $\bar{K}NN-\pi\Sigma{N}$ system manifests itself in the transition 
probabilities on the real energy axis. To investigate the dependence of the resulting transition 
probabilities on models of $\bar{K}N-\pi\Sigma$ interaction, several versions of $\bar{K}N-\pi\Sigma$ 
potentials, which can produce different structures for $\Lambda$(1405) resonance, were used. Within 
this model, we have found a bump produced by $\bar{K}NN-\pi\Sigma{N}$ system appearing in the 
$(Y_{K})_{I=0}+N\rightarrow\pi\Sigma{N}$ transition probabilities in the energy region around the 
$\bar{K}NN-\pi\Sigma{N}$ pole position. We found, that we can find a distinct peak in the mass 
spectrum for momentum $p_{N}=100-250$ MeV/c. In the present calculations, we also found that the 
shape and position of the peaks in the transition probability are independent of the momentum 
$p_{N}$ of the initial $(Y_{K})+N$ channel. Therefore, this fact implies that the bumps are 
corresponding to the $\Lambda$(1405) and $K^{-}pp$ quasi-bound states. Since, the nucleon in the 
initial state covers a continuous, we should include the effect of all momenta in transition 
amplitude. We calculated the averaged transition probability $(Y_{K})_{I=0}+N\rightarrow\pi\Sigma{N}$. 
Furthermore, we have shown that not only we can see the signature of the $K^{-}pp$ quasi-bound state, 
but also, we can see the effect of the branch points in the Faddeev amplitudes (complex plane) and 
transition probabilities which are resulting from $\Lambda$(1405) pole. we have shown that the bump 
structures related to the the branch points can affect the peak corresponding to the $K^{-}pp$ 
quasi-bound state. Thus, in the mesonic decay channel, we should consider the effect of the branch 
points and this reaction would also be helpful to reveal the dynamical origin of $\Lambda$(1405) resonance.

\end{document}